\documentclass{aa}
\usepackage{epsf}

\begin{document}

\thesaurus{07 (08.16.4; 08.15.1; 08.09.2 HS\,2324+3944)}

\title{The photometric behaviour of the peculiar PG\,1159 star HS\,2324+3944
at high frequency resolution
\thanks{Based on observations obtained at the McDonald, Loiano and Beijing
Observatories and at the German-Spanish Astronomical Center, Calar Alto,
operated by the Max-Planck-Institut f\"ur Astronomie Heidelberg jointly with
the Spanish National Commission for Astronomy.}}

\author{R. Silvotti \inst{1} \and S. Dreizler \inst{2} \and G. Handler \inst{3}
\and X.J. Jiang \inst{4}}

\offprints{R. Silvotti (silvotti@na.astro.it)}

\institute{
Osservatorio Astronomico di Capodimonte, via Moiariello 16, I-80131 Napoli,
Italy
\and
Institut f\"{u}r Astronomie und Astrophysik, Waldh\"auser Stra\ss e 64,
D-72076 T\"ubingen, Germany
\and
Institut f\"{u}r Astronomie, Universit\"{a}t Wien, T\"urkenschanzstra\ss e
17, A-1180 Wien, Austria
\and
Beijing Astronomical Observatory and United Laboratory of Optical Astronomy,
CAS, Beijing 100012, China}

\date{Received ......; accepted ...... }

\authorrunning{Silvotti et al.}

\titlerunning{The photometric behaviour of the peculiar PG\,1159 star
HS\,2324+3944 at high freq. resolution}

\maketitle

\begin{abstract}

We present the results from 135 hours of nearly continuous time series
photometry on the ``hybrid'' (H-rich) PG\,1159 variable star HS\,2324+3944,
obtained in August--September 1997.
The power spectrum of the data shows several frequencies (about 20 or more),
concentrated in three narrow and very crowded regions near 475, 390 and
950 $\mu$Hz in decreasing amplitude order.
Most (if not all) of the peaks in the latter region are linear combinations of
the high-amplitude frequencies between 455 and 500 $\mu$Hz.
If we divide the data set into two equal parts, the power spectra are different.
This is probably due to a not sufficiently long (and therefore
not completely resolved) light curve;
nevertheless an alternative hypothesis of a single damped oscillator may not
be completely ruled out.
If we adopt the first hypothesis, the high concentration of peaks between
455 and 500 $\mu$Hz suggests the presence of both l=1 and l=2 high-overtone
nonradial g-modes.
The insufficient frequency resolution of our data does not allow to obtain
definite precision asteroseismology results.
Nevertheless a spacing of the signals is observed, probably due to
stellar rotation with a period of 2.3 days.
If the signal spacing was due to the successive overtones, the period spacings
would be equal to 18.8 (l=1) and 10.4 (l=2) s.


\end{abstract}

\section{Introduction}

\subsection{Pulsating and nonpulsating PG\,1159 stars}

The stars of the PG\,1159 spectral class (31 members) constitute the
intermediate evolutionary phase between the end of the constant luminosity
phase -- at the tip of the asymptotic giant branch (AGB) -- and the beginning
of the white dwarf (WD) cooling phase.
Probing their interior structure provides direct constraints on both classes
of stars and may help to understand better the transition from AGB to WDs and
the nuclear burning turn off process.
A powerful method to probe the interior structure of the pre-WD stars and to
determine some of their basic stellar parameters is given by asteroseismology.
This is possible because 15 PG\,1159 stars and central stars of planetary
nebulae (CSPN) of type [WC], called GW Vir stars from the prototype
(PG\,1159-035), show multiperiodic luminosity variations
which have been interpreted as nonradial g-mode pulsations.
Among them, ten are CSPN, while five appear not
to be surrounded by a nebula (Bradley 1998).
The nature of the luminosity variations of the GW Vir stars was first proven
to be stellar pulsation in the case of PG\,1159-035 itself (Winget et al. 1991).
Nevertheless, and despite the successful results from adiabatic models to
which we will refer to below, the pulsation mechanism of the GW Vir stars is
still not well understood.
Although almost all authors agree that the pulsations should be driven by the
$\kappa$--$\gamma$ mechanism, based on the C/O cyclic ionization (Starrfield et
al. 1984), a good agreement between spectroscopic abundances and observed
pulsation periods has not been yet found (Bradley \& Dziembowski 1996).
A new element in this picture was recently added by Dreizler \& Heber (1998):
their results suggest that the GW Vir pulsations could be related to the
nitrogen abundance.
On the other hand, in the adiabatic pulsation field, theory may explain several
observed phenomena as frequency splitting due to rotation and/or magnetic
fields, period spacing of successive overtones, and variations around the
average period spacing caused by mode trapping in the outer layers of the
star (Kawaler \& Bradley 1994 and references therein).
However, the measurement of frequency and period spacing, which leads to 
accurate determination of rotation, weak magnetic fields, stellar mass,
external layer masses, and even luminosity and distance, needs power spectra with low noise
and high frequency resolution.
For these reasons it is necessary to obtain long and nearly continuous data
sets, such as those obtained by the Whole Earth Telescope network
(Nather et al. 1990).

\subsection{HS\,2324+3944}

The star HS\,2324+3944 (hereafter HS\,2324) is one out of four peculiar
members of the PG\,1159 spectral class showing strong H Balmer absorption in
their spectra (Dreizler et al. 1996), called ``hybrid PG\,1159 stars''
(Napiwotzki \& Sch\"onberner 1991) or lgEH\,PG\,1159, following the notation
scheme of Werner (1992).
It has an effective temperature of (130\,000\,$\pm$\,10\,000)\,K and a
surface gravity $\log\,g$=6.2\,$\pm$\,0.2 (Dreizler et al. 1996).
Recent new analysis of the HST-GHRS spectrum of HS\,2324 show that the C/He
and O/He ratio (0.4 and 0.04 by number) is as high as in ordinary PG\,1159
stars (Dreizler 1998).
Therefore only the hydrogen abundance (H/He=2 by number) makes it unusual.
HS\,2324 does not show direct signs of on-going mass loss from P\,Cygni shaped
line profiles, like several other luminous PG\,1159 stars (Koesterke et al.
1998). However, detailed line profiles from high resolution Keck spectroscopy
show evidence of mass loss in the order of roughly $10^{-8}\,M_{\odot}/yr$
(Dreizler et al., in preparation).
Regarding effective temperature and gravity, it belongs to the subgroup of
luminous PG\,1159 stars which are in general Central Stars of Planetary
Nebulae.
However, differently from all the other known hybrid PG\,1159 stars, no nebula
is detected around HS\,2324 (Werner et al. 1997).

HS\,2324 was discovered to be variable by Silvotti (1996).
Handler et al. (1997), with more extensive observations, showed that at least
four different frequencies were active and therefore that the
GW Vir hypothesis was the most likely.
For other two hybrid PG\,1159 stars, the nuclei of A\,43 and NGC\,7094,
periodic light variations are only suspected (Ciardullo \& Bond 1996).
The interest for the variability of HS\,2324 is enhanced by its hydrogen
abundance.
The presence of H was generally considered as a inhibitor of pulsations
(Stanghellini et al. 1991).
First steps to test the effects of the presence of H in the driving
regions have been undertaken by Saio (1996) and Gautschy (1997).
The models of Saio (1996) do pulsate with 3$\%$ of H mass fraction.
The models of Gautschy (1997) are able to reproduce the observed periods of
HS\,2324, with a very similar H abundance of 20$\%$ by mass.

For all the reasons stated above, HS\,2324 is a very interesting star:
the analysis of its photometric behaviour at high frequency resolution
may give important results not only for a detailed study of the star itself,
but also for more general questions regarding the GW Vir pulsation phenomenon.
Therefore we decided to carry out a multisite photometric campaign on HS\,2324,
which may be considered as a first step for successive more extensive
campaigns.

\section{Observations}

The multisite campaign on HS\,2324 was performed during 2 weeks in
August--September 1997, centered on new Moon.
The journal of observations in Table 1 gives information on the observatories
involved, telescopes and instruments used, and duration of the single runs.
Most observations were obtained using two or three channel photometers
with bialkali photomultipliers (EMI9784QB for Loiano, Hamamatsu R647 for
Beijing and McDonald), no filters, and an integration time of 10\,s,
which was subsequently merged to 90\,s.
The leak of sensitivity to periods shorter than 180\,s did not give us any
trouble because no signals were detected in that range from a preliminary
analysis of the photometer data alone.
Only the Calar Alto data were collected using a SITe\#1d CCD, B filter, and
an exposure time of about 20\,s (first two nights) or 40\,s (following three
nights) for each datum; the times between successive data points vary between
about 70 and 100\,s.
The error introduced by the different effective wavelengths of each detector
has been evaluated to be not more than 5$\%$ in amplitude.

Despite the small number of participants in the campaign, only four,
we obtained a good coverage, comparable with other multisite campaigns,
thanks to the good weather conditions in most nights at the different sites.
The complete (and combined) light curve is shown in Figure 1; it has a total
duration of 134.9 hours, with an overall duty cycle of 43$\%$.
In the central part of the run (40.8 hours), the duty cycle is 98$\%$.

\begin{table*}[ht]
\begin{center}
\caption[]{Journal of the observations}
\begin{tabular}{lccccc}
\hline
\hline
\multicolumn{1}{c}{\bf Telescope} &
\multicolumn{1}{c}{\bf Instrument} &
\multicolumn{1}{c}{\bf Observer} &
\multicolumn{1}{c}{\bf Date} &
\multicolumn{1}{c}{\bf Start Time} &
\multicolumn{1}{c}{\bf Run Length} \\
 & & &
\multicolumn{1}{c}{\bf (UT)} &
\multicolumn{1}{c}{\bf (UT)} &
\multicolumn{1}{c}{\bf (hours)} \\
\hline
\hline
McDonald 2.1 m   & PMT & GH & 26 Aug 1997 & 07:31:20 & 3.55 \\
McDonald 2.1 m   & PMT & GH & 27 Aug 1997 & 04:45:00 & 6.33 \\
Loiano 1.5 m     & PMT & RS & 27 Aug 1997 & 20:19:16 & 3.54 \\
McDonald 2.1 m   & PMT & GH & 28 Aug 1997 & 03:16:00 & 7.66 \\
McDonald 2.1 m   & PMT & GH & 29 Aug 1997 & 03:10:30 & 7.59 \\
Loiano 1.5 m     & PMT & RS & 29 Aug 1997 & 23:32:57 & 1.00 \\
Loiano 1.5 m     & PMT & RS & 30 Aug 1997 & 21:57:57 & 4.84 \\
McDonald 0.9 m   & PMT & GH & 31 Aug 1997 & 03:49:30 & 0.58 \\
Loiano 1.5 m     & PMT & RS & 31 Aug 1997 & 22:01:01 & 5.00 \\
Beijing 0.85 m   & PMT & JX & 01 Sep 1997 & 16:17:20 & 1.03 \\
Calar Alto 2.2 m & CCD & SD & 01 Sep 1997 & 19:58:51 & 8.98 \\
Loiano 1.5 m     & PMT & RS & 01 Sep 1997 & 21:30:01 & 5.49 \\
Beijing 0.85 m   & PMT & JX & 02 Sep 1997 & 11:59:40 & 8.39 \\
Loiano 1.5 m     & PMT & RS & 02 Sep 1997 & 20:05:19 & 4.29 \\
Calar Alto 2.2 m & CCD & SD & 02 Sep 1997 & 22:16:23 & 6.20 \\
McDonald 0.9 m   & PMT & GH & 03 Sep 1997 & 02:39:40 & 8.73 \\
Beijing 0.85 m   & PMT & JX & 03 Sep 1997 & 12:04:20 & 8.27 \\
Calar Alto 2.2 m & CCD & SD & 03 Sep 1997 & 20:30:41 & 8.29 \\
Beijing 0.85 m   & PMT & JX & 04 Sep 1997 & 16:03:30 & 4.10 \\
Loiano 1.5 m     & PMT & RS & 04 Sep 1997 & 19:23:35 & 5.77 \\
Calar Alto 2.2 m & CCD & SD & 04 Sep 1997 & 19:32:06 & 9.28 \\
McDonald 0.9 m   & PMT & GH & 05 Sep 1997 & 05:31:30 & 1.76 \\
Calar Alto 2.2 m & CCD & SD & 05 Sep 1997 & 19:32:18 & 8.82 \\
McDonald 0.9 m   & PMT & GH & 06 Sep 1997 & 02:39:10 & 8.77 \\
McDonald 0.9 m   & PMT & GH & 07 Sep 1997 & 06:05:00 & 5.38 \\
McDonald 0.9 m   & PMT & GH & 08 Sep 1997 & 02:25:40 & 9.10 \\
\hline
\end{tabular}
\end{center}
\end{table*}

\begin{figure*}[ht]
\vspace{176mm}
\includegraphics{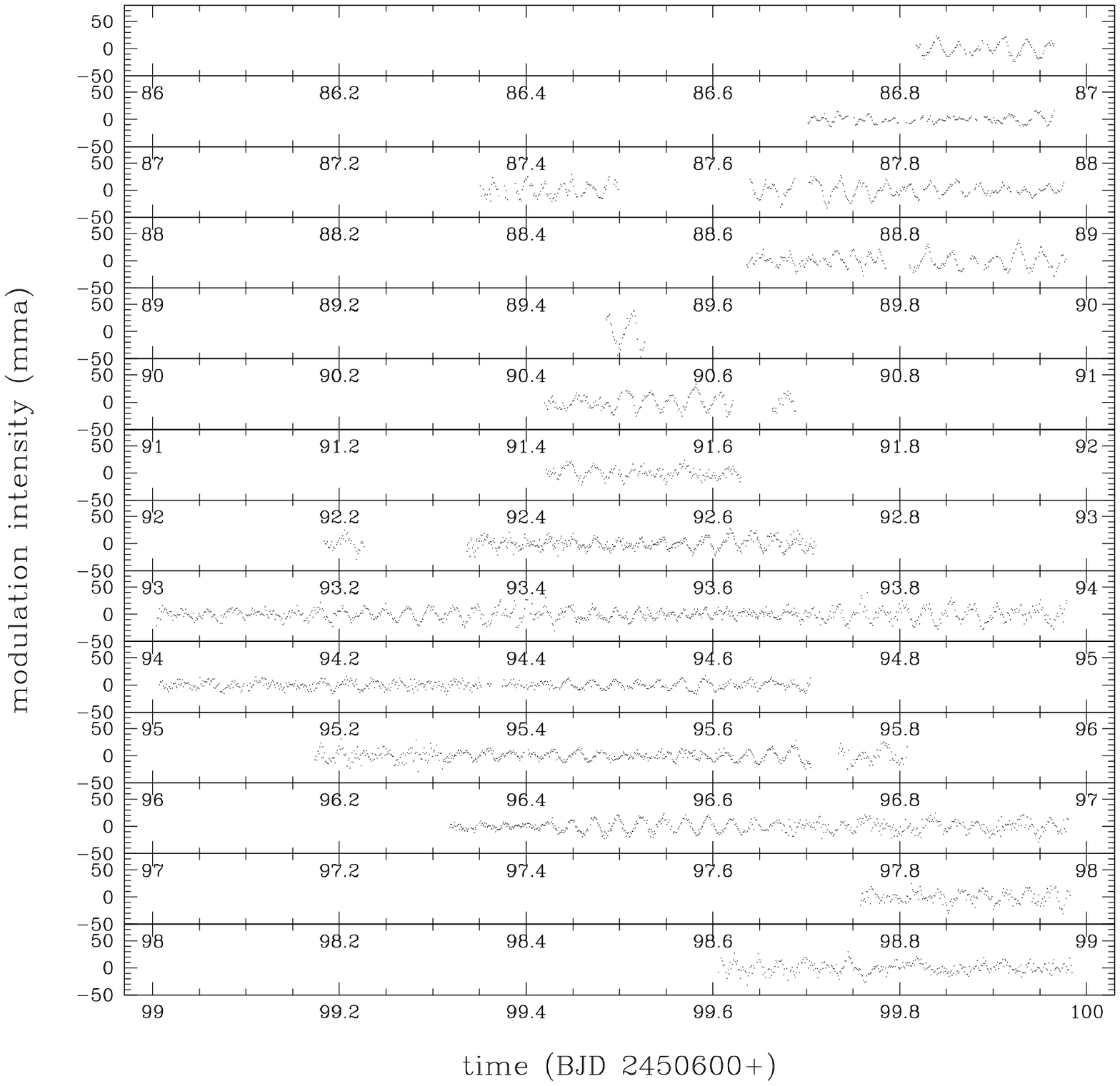}
\caption[]{The complete light curve of HS\,2324; each panel represents 24
hours.}
\end{figure*}

\subsection{Details on data acquisition and reduction}

We followed basically the same data reduction procedure as described in
Handler et al. (1997). Here we summarize this procedure and give some detail 
on a few differences.

For the Beijing and McDonald photoelectric data, we chose the same comparison
star already used by Handler et al. (1997), which was also one of the
comparison stars used in the Calar Alto CCD measurements.
For the Loiano photoelectric data this was not possible because the box of
channel 2 is more distant from channel 1: therefore we used the same comparison
star already used by Silvotti (1996).
Both comparison stars were tested again for photometric constancy and found 
not to be variable.
We then turned to sky subtraction. For the Beijing measurements, where a
third channel was available, the sky background could be monitored
simultaneously. In this case we subtracted the sky counts on a point by point
basis. To reduce the scatter of the background measurements, some smoothing
was applied whenever possible.
At McDonald and Loiano only two channels were available.
In this case sky was measured using channel 1 and 2 for about 1 min at
irregular intervals of typically 20--90 min, depending on sky stability and
presence of the Moon.
The sky counts were then interpolated linearly and subtracted.
In a few cases we used a cubic spline for the sky interpolation,
when it was clear that this procedure was giving better results than the
linear fit.
All the PMT data were then corrected for extinction. Afterwards, they were
used to examine possible transparency variations.
In a few cases of high sky instability, the count ratio between channel 1
and channel 2 was used instead of channel 1 counts only. Some smoothing
of the channel 2 data was applied when possible.
Systematic long time scale trends ($>$ 2 hours), probably due to tube drifts
and/or to residual extinction, were finally compensated by means of linear or
cubic spline interpolation.

For the Calar Alto CCD data, 10 comparison stars were selected, after having
been tested for photometric constancy.
Their average magnitude was subtracted from the HS\,2324 measurements on a
point by point basis.
Differential extinction was corrected by means of a cubic spline.

Finally all the single data sets (PMT + CCD) were set to a mean value of zero.
The times of all data were then converted to Barycentric Julian Date using
the algorithm of Stumpff (1980).
The accuracy of the original times was of the order of $\pm$ 0.3\,s (Beijing),
$\pm$ 0.01\,s (Loiano), $\pm$ 0.2\,s (McDonald) and $\pm$ 1\,s (Calar Alto).
For Calar Alto 1\,s is also the time accuracy of each measurement.
Moreover, to have a more homogeneous data set, we have binned the PMT data
to an effective integration time of 90\,s.
The value of 90\,s has been chosen because it corresponds to the mean
distance between consecutive CCD observations.
When more than one site was active at the same time, in the overlap regions,
we applied a weighted average of the data obtained at the different sites.
In this way even lower quality data can be used to improve the S/N ratio (see
Moskalik 1993).
In its final form, the data set is constituted by the time of each integration,
the fractional departure of the count rate from the mean (modulation
intensity), and the error.

\section{Temporal spectroscopy}

\subsection{Spectral window and power spectrum}

\begin{figure*}[htb]
\vspace{118mm}
\includegraphics{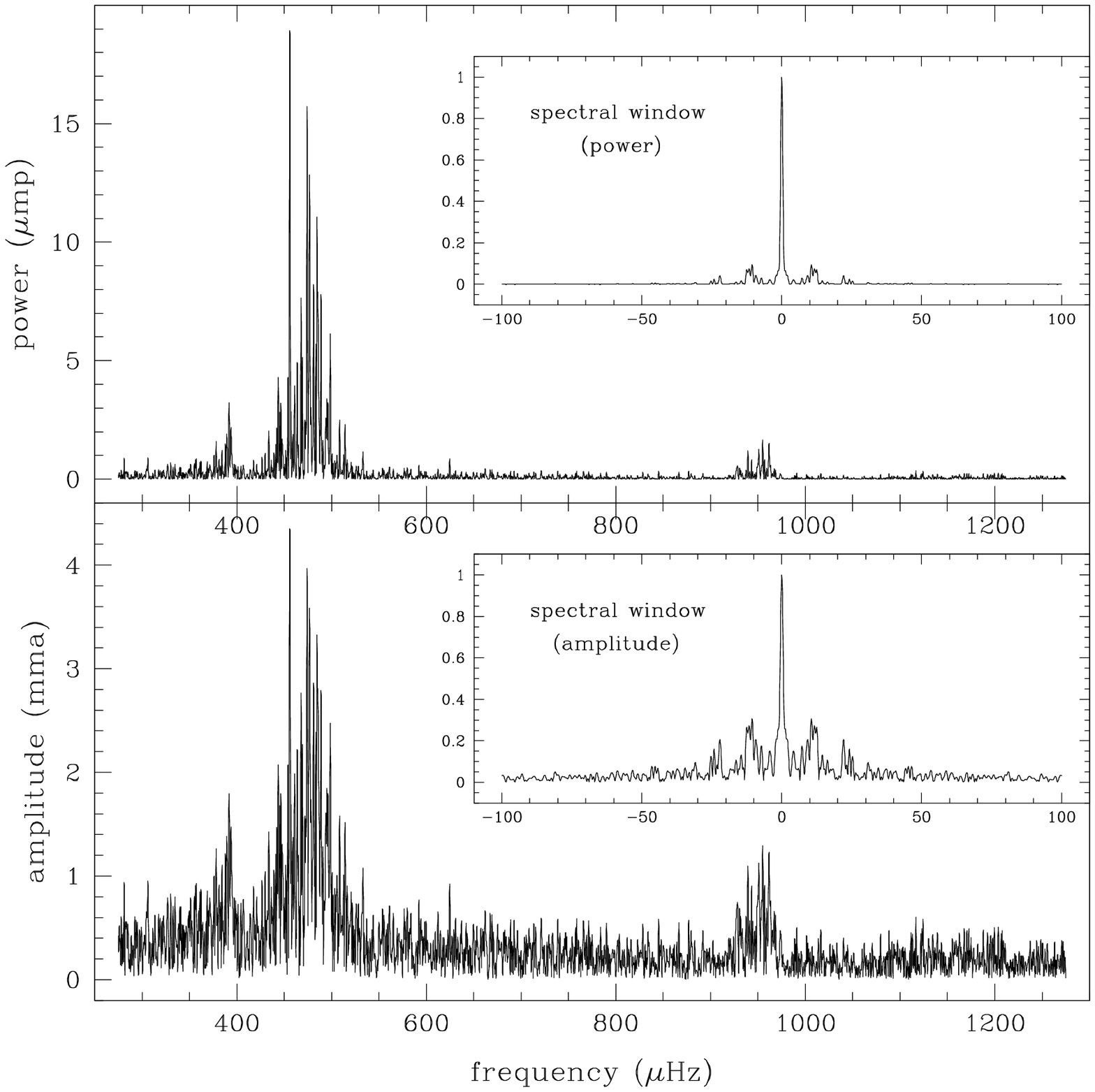}
\caption[]{Power spectrum (top) and amplitude spectrum (bottom) of the entire
data set. The DFT of the window function (spectral window) is also reported
both in power and in amplitude; we can note that the 1 cycle/day sidelobes
have maxima at $\pm$ 10.6 $\mu$Hz, corresponding to a period of 1.09 days.}
\end{figure*}

\begin{figure*}[htb]
\vspace{176mm}
\includegraphics{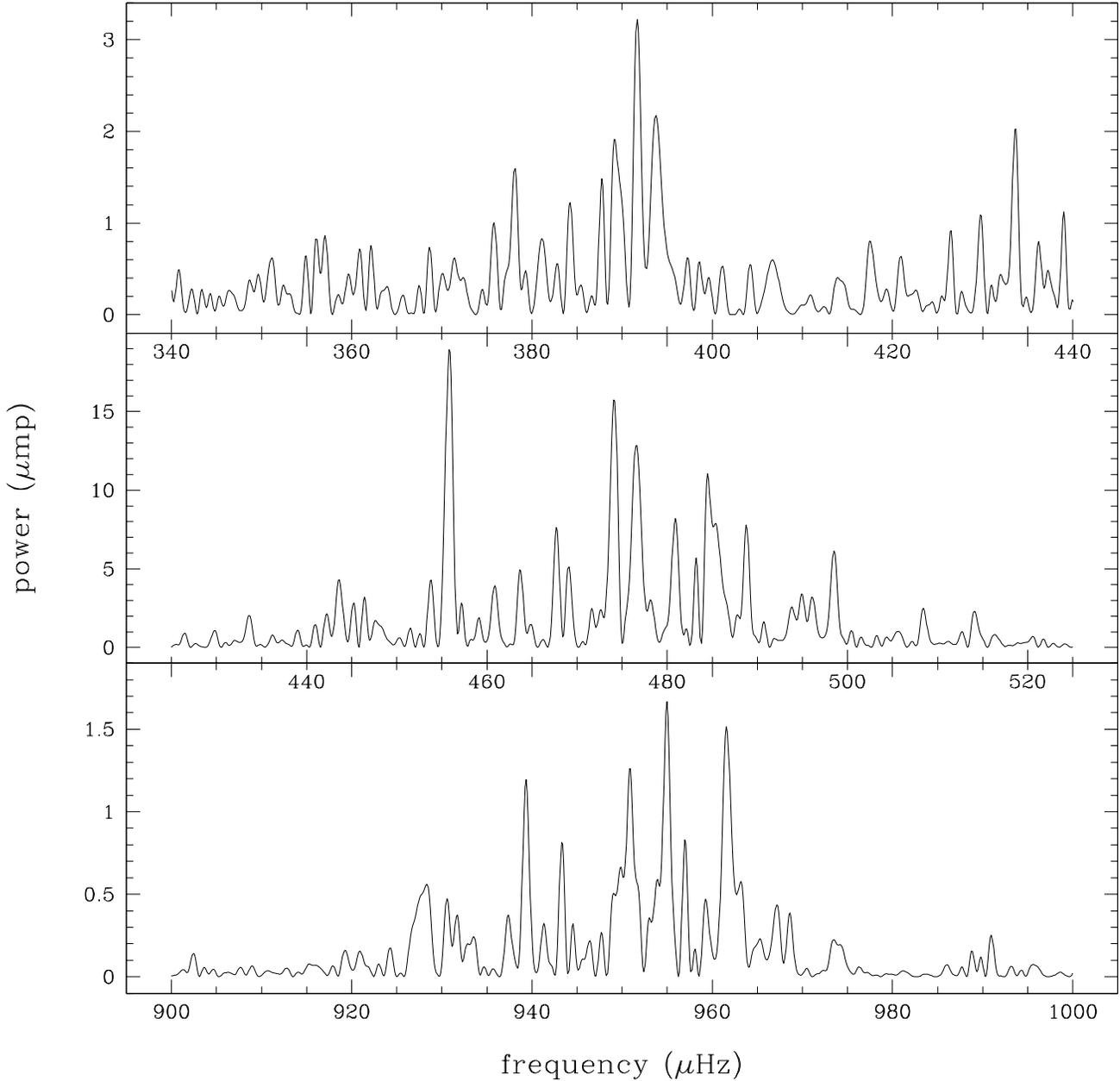}
\caption[]{Detailed power spectrum of the entire data set.
Note that different vertical scales have been used.}
\end{figure*}

\begin{figure*}[htb]
\vspace{118mm}
\includegraphics{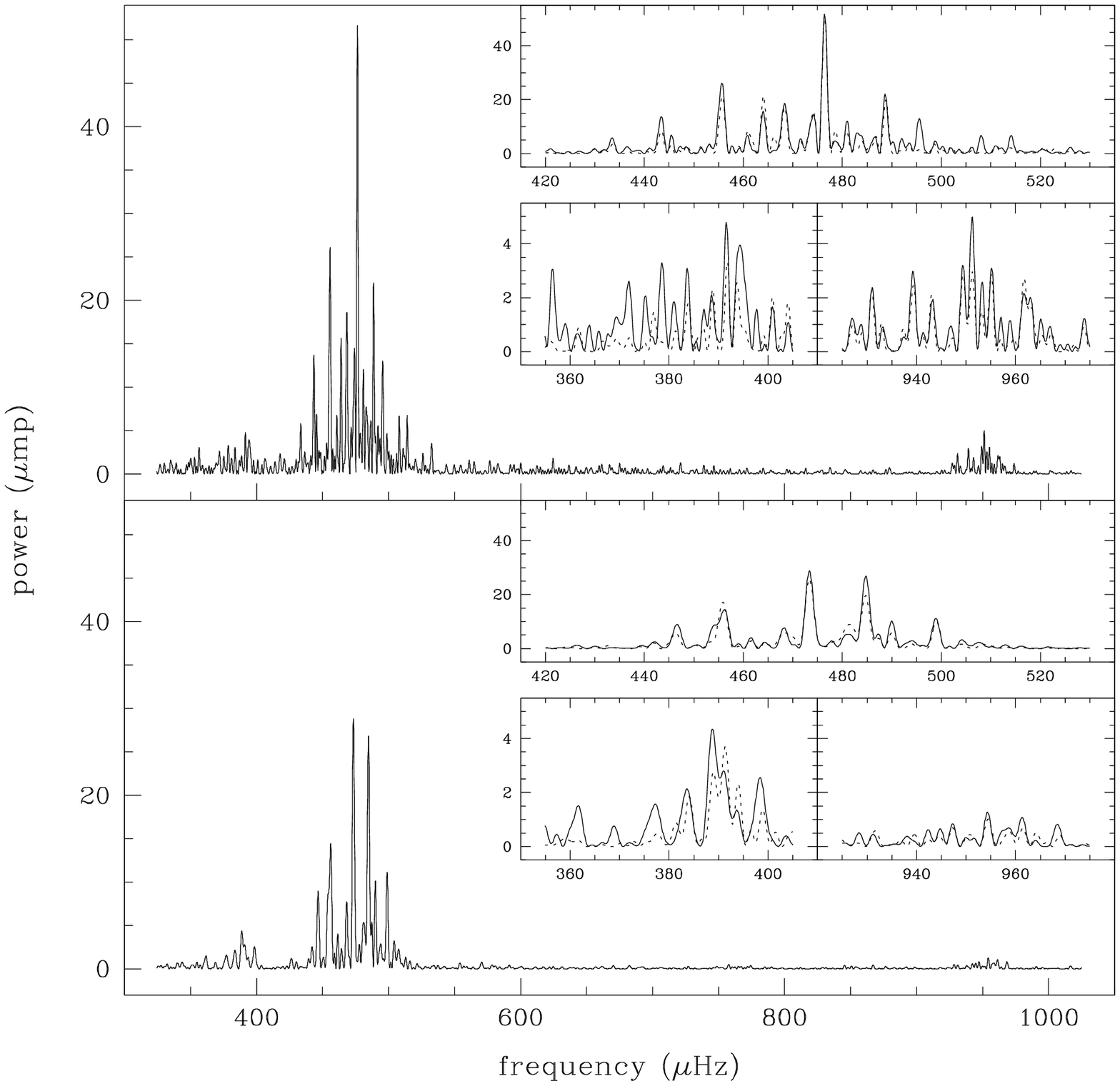}
\caption[]{Power spectrum of the first half of data (top) versus second half
(bottom). Note the strong differences both in power and in frequency.
In the small panels the regions of interest are highlighted;
the dotted lines represent the spectra of the first and second half of a
synthetic light curve obtained summing the 19 sinusoids listed in Table 2 and
having same time sampling as the HS\,2324 data.
No noise was added to the synthetic light curves.}
\end{figure*}

We computed a single sine function with unit amplitude at the same sample times
of the entire data set (window function).
The discrete Fourier transform of the window function gives the spectral
window, which is shown in Figure 2.
For more completeness, both the amplitude and the power (amplitude squared)
spectra of the window function are presented.
The only structures which may give troubles for the unambiguous identification
of the modes are the 1 cycle/day aliases, with a relative amplitude of about
0.3 (relative power of 0.09).

Using the same discrete Fourier transform (DFT), based on the Deeming (1975)
method and Kurtz (1985) algorithm, we computed the transform of the entire
reduced data set.
In Figure 2 both amplitude and power spectrum are shown in units of
millimodulation amplitude (mma) and micromodulation power ($\mu$mp), following
the suggestion of Winget et al. (1994).
We clearly see that the signals are concentrated in three main regions near
390, 470 and 950 $\mu$Hz. These regions are highlighted in Figure 3.
The tested frequency resolution of our data set is 1.4 $\mu$Hz, according to
Loumos \& Deeming (1978);
such a value corresponds to obtain a half amplitude separation of two close
peaks with equal amplitude.

Looking at Figure 3, some structures appear to be not completely resolved,
suggesting that the light curve could be too short that we can resolve all the
present frequencies.
To test this crucial point, we divided the data set into two equal parts and
computed the Fourier transform of each part.
The two power spectra, presented in Figure 4, show strong differences not only
in amplitude, but also in frequency.
The first interpretation of such differences is simply that the data set is
not long enough to ``stabilize'' the Fourier transform.
In other words, the light curve is not completely resolved.
The reliability of this hypothesis is increased by the fact that our best
multisinusoidal fit of the entire data set (see next section and Table 2)
gives good results also when applied only to the first or the second
half of data (Figure 4).
If the DFT apparent instability is actually due to the insufficient coverage,
most of the analyses reported in sections 3.2, 4.1 and 4.2, and based on the
assumption that the DFT of HS\,2324 is not time dependent on time scales
shorter than our run, will need further confirmation from a new longer
observational campaign.

On the other hand, if the DFT time instability was real, we would need a
different explanation for such peculiar behaviour.
An alternative hypothesis of a fast damped oscillator has been considered
and is reported in Section 5.

\subsection{Frequency identification}

Looking at Figure 2 and 3, it is immediately evident that determining
the active frequencies from the power spectrum of HS\,2324 will be more
difficult than in most other GW Vir stars for several reasons:
the power is concentrated in only 3 crowded regions;
the amplitudes are very low;
the low frequencies imply that the frequency and the period spacing expected
may have about same values.
The high frequency region does not help much because it seems to be
constituted only by linear combinations of the low-frequency peaks.
Moreover we know that the frequencies are not completely resolved and
therefore we certainly have errors both in frequency and in amplitude.

To distinguish the real frequencies present in the HS\,2324 data from the
artifacts introduced by the spectral window, we proceeded as follows.
First we selected the highest peak in each of the three ``active regions''
near 390, 470 and 950 $\mu$Hz.
The separation of the three active regions guarantees that the aliases
of each frequency have almost zero influence in the other two regions.
Second we applied a least-squares multisinusoidal fit to the data to
determine accurate amplitudes and phases of the three selected sine waves.
Third we created an artificial signal adding together the three sinusoids
and using the same sampling times as the data.
This artificial signal was then subtracted from the data (prewhitening), and
the residuals were analyzed again.
Three (or less) new frequencies were selected and the whole procedure was
repeated n times until the power of the prewhitened data was near the level of
the noise.
At each iteration we selected first those frequencies which were not
coincident with the one day aliases of the strongest signals.
At the end of the whole process, the frequencies, amplitudes and phases
were optimized with a final least-squares fit with all the frequencies found.
The resulting best fit parameters are listed in Table 2.
It is important to emphasize, however, that the solution in Table 2 is not the
only one.
After having performed the prewhitening of 7 frequencies (marked with an
asterisk in Table 2), different solutions become possible.
The frequencies selected in Table 2 represent the result of several attempts.
The solution that we have chosen has the advantage that it produces small
residuals with a relative small number of frequencies (Figure 5).
Looking at Table 2, we can note that most (if not all) of the high frequency
signals correspond to linear combinations of the high-amplitude frequencies.
\begin{table*}[ht]
\begin{center}
\caption[]{Results of the sinusoidal fit}
\begin{tabular}{rcccc}
\hline
\hline
\multicolumn{1}{c}{\bf Frequency} &
\multicolumn{1}{c}{\bf Period} &
\multicolumn{1}{c}{\bf Amplitude} &
\multicolumn{1}{c}{\bf T$_{MAX}^{1}$} &
\multicolumn{1}{c}{\bf Comments} \\
\multicolumn{1}{c}{\bf ($\mu$Hz)} &
\multicolumn{1}{c}{\bf (s)} &
\multicolumn{1}{c}{\bf (mma)} &
\multicolumn{1}{c}{\bf (BJD 2450686.+)} &
 \\
\hline
\hline
     389.27$\pm$0.06 & 2568.89$\pm$0.41 & 1.33$\pm$0.15 & 0.82337$\pm$0.00145 &
      f1 \\
$^*$ 391.64$\pm$0.06 & 2553.38$\pm$0.37 & 1.50$\pm$0.15 & 0.84665$\pm$0.00135 &
      f2 \\
     393.51$\pm$0.06 & 2541.25$\pm$0.37 & 1.45$\pm$0.15 & 0.82707$\pm$0.00133 &
      f3 \\
$^*$ 455.74$\pm$0.02 & 2194.23$\pm$0.10 & 4.27$\pm$0.15 & 0.83075$\pm$0.00041 &
      f4 \\
     460.68$\pm$0.06 & 2170.70$\pm$0.27 & 1.84$\pm$0.16 & 0.83464$\pm$0.00117 &
      f5 \\
     472.90$\pm$0.05 & 2114.63$\pm$0.23 & 3.28$\pm$0.23 & 0.82214$\pm$0.00098 &
      f6 \\
$^*$ 473.87$\pm$0.05 & 2110.29$\pm$0.24 & 3.29$\pm$0.22 & 0.82515$\pm$0.00097 &
      f7 \\
     476.13$\pm$0.04 & 2100.25$\pm$0.19 & 4.69$\pm$0.45 & 0.84137$\pm$0.00089 &
      f8 \\
$^*$ 476.64$\pm$0.04 & 2098.02$\pm$0.17 & 5.20$\pm$0.45 & 0.84039$\pm$0.00078 &
      f9 \\
     480.95$\pm$0.04 & 2079.22$\pm$0.18 & 2.25$\pm$0.15 & 0.83641$\pm$0.00077 &
      f10 \\
     485.11$\pm$0.03 & 2061.37$\pm$0.14 & 3.13$\pm$0.17 & 0.83581$\pm$0.00061 &
      f11 \\
     498.78$\pm$0.04 & 2004.91$\pm$0.18 & 1.98$\pm$0.15 & 0.82382$\pm$0.00082 &
      f12 \\
     930.49$\pm$0.11 & 1074.71$\pm$0.13 & 1.02$\pm$0.21 & 0.82086$\pm$0.00109 &
      (f4+f7) \\
     931.71$\pm$0.12 & 1073.29$\pm$0.13 & 0.94$\pm$0.20 & 0.82618$\pm$0.00111 &
      f4+f8 (f4+f9) \\
$^*$ 939.42$\pm$0.11 & 1064.49$\pm$0.13 & 1.08$\pm$0.21 & 0.81967$\pm$0.00105 &
      (f4+f11)\\
     949.25$\pm$0.16 & 1053.46$\pm$0.17 & 0.76$\pm$0.22 & 0.82672$\pm$0.00144 &
      f6+f8 (f6+f9) \\
$^*$ 955.04$\pm$0.10 & 1047.07$\pm$0.11 & 1.22$\pm$0.20 & 0.82392$\pm$0.00091 &
      f7+f10 (f4+f12) \\
$^*$ 961.71$\pm$0.12 & 1039.82$\pm$0.13 & 1.04$\pm$0.22 & 0.81880$\pm$0.00120 &
      f9+f11 (f10+f10) \\
     963.15$\pm$0.13 & 1038.26$\pm$0.14 & 0.96$\pm$0.21 & 0.82792$\pm$0.00120 &
      \\
\hline
\end{tabular}
\end{center}
Notes: $^{(1)}$ Time of the first maximum inside the data set.
$^{(*)}$ These frequencies are the most reliable (see the text).

\end{table*}

\begin{figure*}[htb]
\vspace{118mm}
\includegraphics{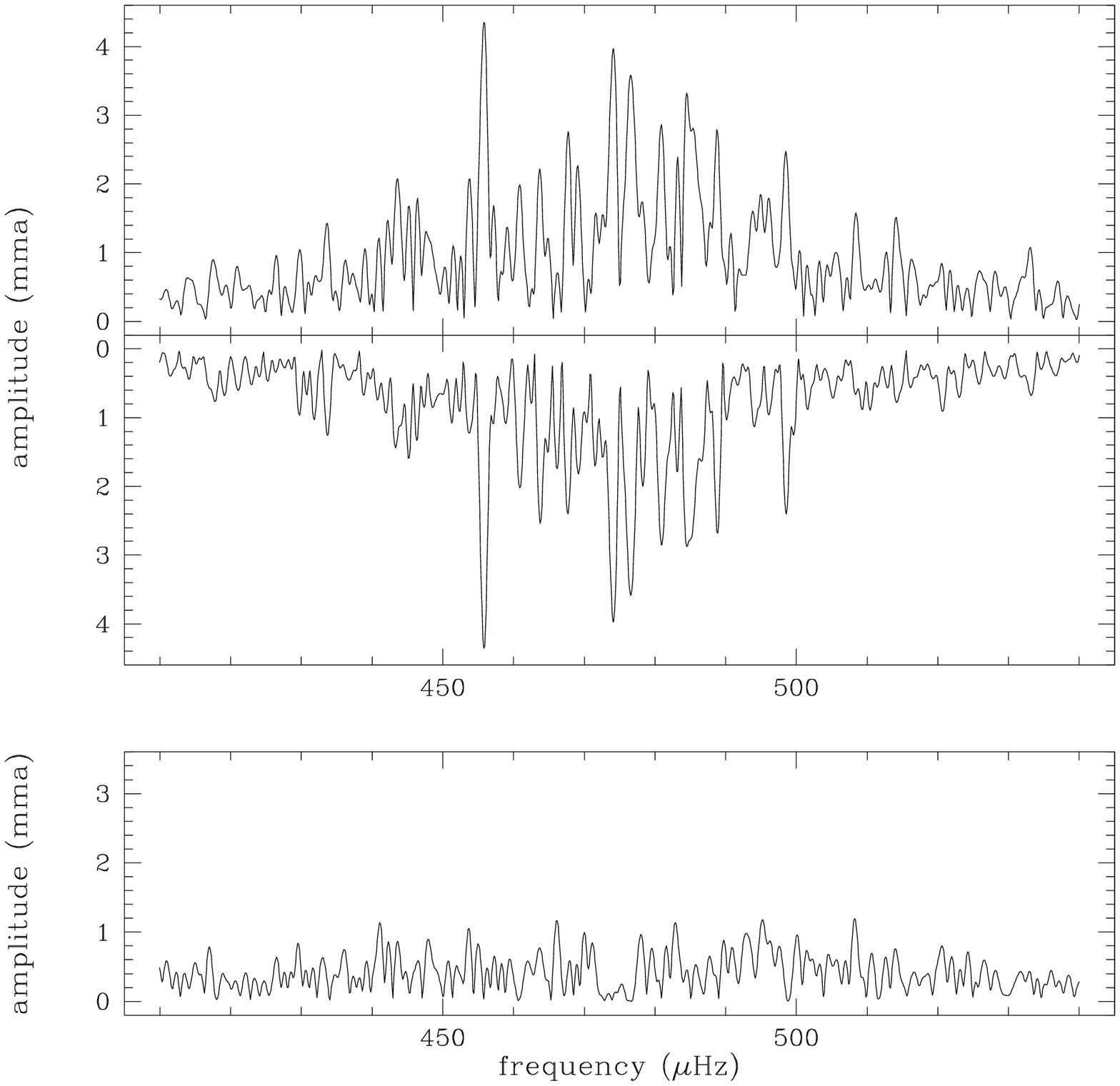}
\caption[]{Upper panels: amplitude spectrum of the entire data set (top)
compared with the spectrum of the 19-frequency fit (bottom) having the same
time sampling as the data. No noise was added to the synthetic data.
Lower panel: spectrum of the residuals.
Note that the vertical scale is the same in all panels.}
\end{figure*}

\section{``Classical'' seismological interpretation}

\subsection{Frequency splitting}

In the power spectrum of HS\,2324 there is not any clear direct sign of
frequency splitting. This may be due to different reasons.
A very small rotation rate, with secondary (m$\ne$0) modes below the frequency
resolution, seems quite unlikely because it would require a rotation
period longer than about 9 days.
A more realistic possibility is that the star has a low inclination,
so that the amplitudes of the m$\ne$0 modes are near the level of the noise.
A third possibility is that the concentration of the peaks is so high that
we are simply not able to recognize the modes splitted by the rotation.

If a direct identification of the rotational splitting is not possible,
the high number of peaks allows one to make use of statistical methods.
First we constructed an histogram of the frequency separations between the
signals listed in Table 2, excluding the linear combinations.
The result was not significant: no preferred frequency spacings appeared.
Another attempt was done using all the peaks of the power spectrum higher
than a fixed level; we selected 49 frequencies and made the histogram.
Here also we did not get any significant result apart that all the peaks were
never higher than the one day alias near 10--11 $\mu$Hz.

At this point we invoked a third method: we computed the DFT of the
amplitude spectrum, using two different subsets spanning 350--1000 $\mu$Hz
(all signals) and 440--500 $\mu$Hz (only high power signals).
The resulting power spectra are shown in Figure 6 (left panels).
Considering the lower panel, the highest peaks are at 10.1 and 4.2 $\mu$Hz;
a third peak at 2.5 $\mu$Hz is clearly visible and more evident in the upper
panel.
Let us focus our attention to the latter two (for the first one we will
give an interpretation below): their ratio, equal to 0.594 (or 0.589),
is very close (1\% level) to the canonical value of 0.6 predicted
by asymptotic theory for the ratio between l=1 and l=2 rotational frequency
splitting.
Therefore the 4.2 and 2.5 $\mu$Hz peaks might correspond to the frequency
separation between m and m$\pm$1 (l=2 and l=1) modes.
The corresponding rotation period of the star would be $P_{ROT}=2.31\pm0.15$
days.
This result may not be considered definitive because the method used is very
sensitive to noise.
Moreover the signal that we are looking for is not actually coherent:
the constant frequency separation between the modes of the same overtone
splitted by the rotation does not correspond, in general, to the separation
between successive overtones, which is not constant in frequency
(but almost constant in our particular case, due to the narrowness of the
high-power region).
From this point of view a more appropriate -- but not much less noisy -- method
to measure the frequency spacing is given by the autocorrelation of the DFT
(Press et al. 1992).
The results of the DFT autocorrelation, reported in Figure 6 (right panel),
are less significant than, but do not contradict, those obtained from the DFT
of the amplitude spectrum.

Looking now at the peak near 10 $\mu$Hz of Figure 6, its frequency
separation is very close to that of the one day alias;
therefore we could conclude that it is actually produced by all the
aliases of the signals.
This conclusion would give more confidence in the rotational origin of the two
peaks at 4.2 and 2.5 $\mu$Hz. Moreover comparing the power of these two
peaks with that of the one day alias, we could suppose that the weakness
of the m$\ne$0 modes is actually due to the low inclination of the star.
But with a deeper analysis (testing the variations of the three peaks
of Figure 6 (left panels) when we subtract different signals from the HS\,2324
data (prewhitening)), we can easily demonstrate that the peak at about
10 $\mu$Hz has at least two components: one at 10.6 $\mu$Hz actually
related to the one day alias and another one related to the separation
between the two signals at about 474 and 485 $\mu$Hz in the data DFT.
Therefore it is more difficult to derive any consideration about the weakness
of the m$\ne$0 modes and the low inclination hypothesis does not have any
support.
On the other hand, we can also demonstrate that the origin of the 2.5 and
4.2 $\mu$Hz peaks is strongly related to the separation between a few
large amplitude signals.
Conclusion: if the frequency spacings of 2.5 and 4.2 $\mu$Hz are actually
due to the stellar rotation, the low inclination hypothesis can not be longer
followed. The new even more simple picture would be the following:
there are five l=1 triplet component candidates (474.1
\footnote{The frequencies reported here are not taken from Table 2;
we prefer to use the values found in the data DFT, as Table 2 may
contain errors.}
\hspace{-1.5mm}
and 476.6 $\mu$Hz plus 389.1, 391.7 and 393.8 $\mu$Hz) and there are three
l=2 quintuplet component candidates (480.9, 484.5 and 488.8 $\mu$Hz).
If we derive the frequency spacing from these values we obtain a rotation
period of the star $P_{ROT}=2.41\pm0.22$ days, slightly different from the
previous one.
Other possible multiplets might be present at 483.2 and 485.4 $\mu$Hz (l=1),
and 463.7 and 467.7 $\mu$Hz (l=2).

In this context we can also try to estimate the inclination of the star using
the l=1 modes.
Following Pesnell (1985)
\footnote{These equations require two strong assumptions, certainly not
completely -- if not at all -- realistic for low gravity GW Vir stars:
the m$\ne$0 modes should be all excited at the same amplitude level;
the amplitude variations during the run must be excluded.
Therefore this estimate of the inclination of the stellar rotational axis must
be considered very tentative.}
\hspace{-1.5mm}
we obtain an indication for $i\simeq50^{\circ}$.

\begin{figure*}[htb]
\vspace{88mm}
\includegraphics{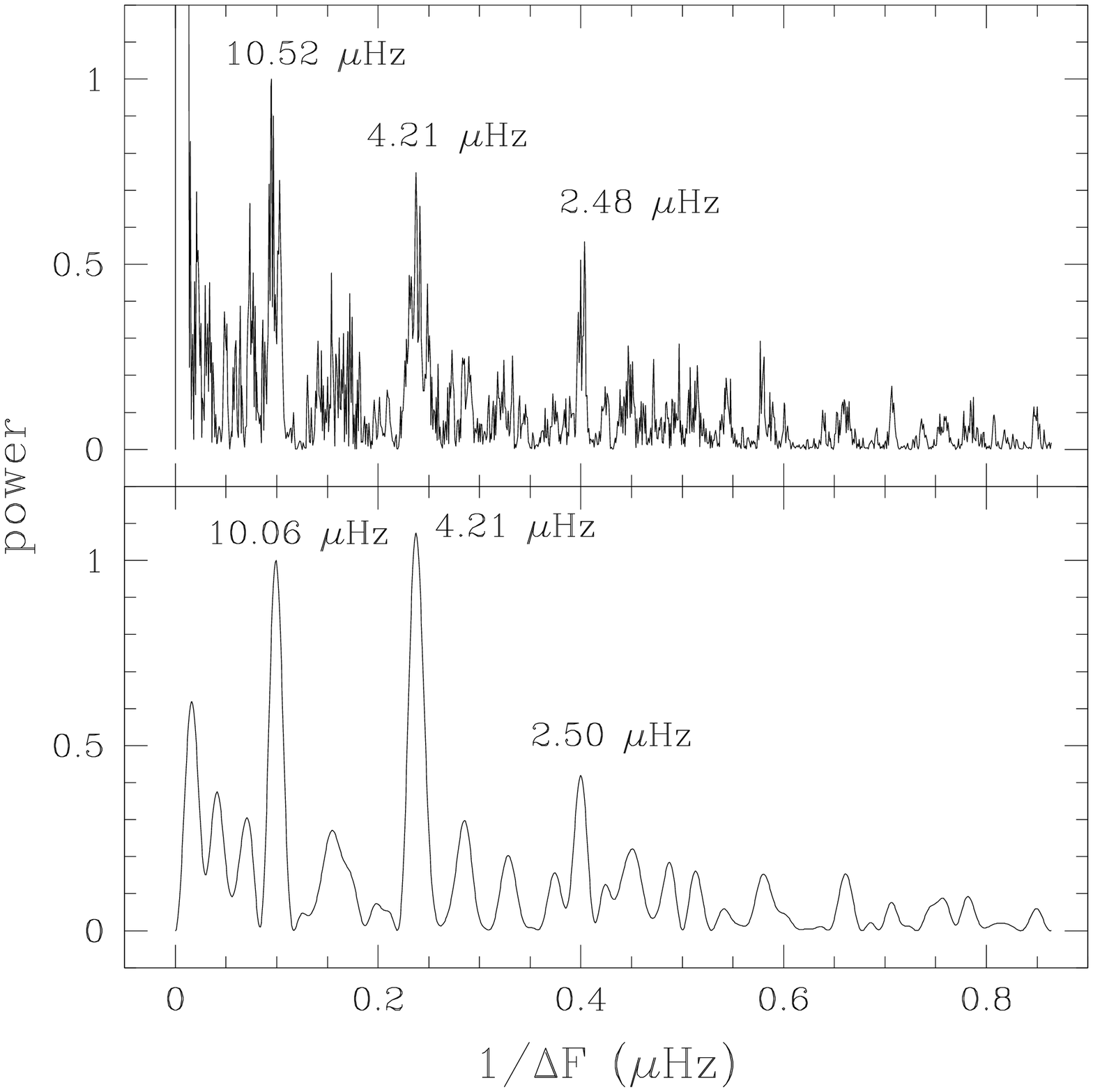}
\includegraphics{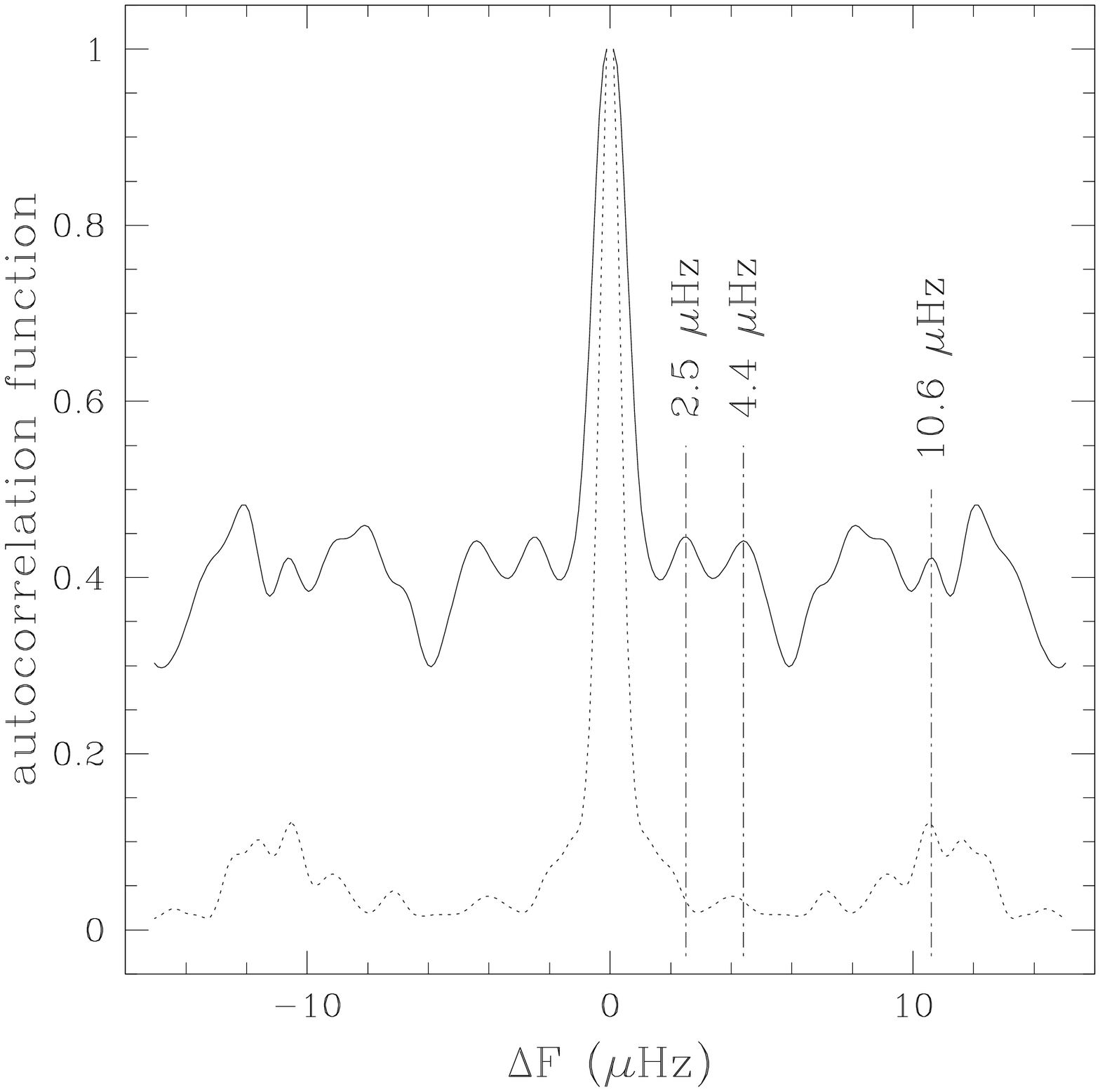}
\caption[]{Search for the frequency spacing.\\
Left panels: power spectrum of the data amplitude spectrum using two different
subsets of the DFT: 350\,$\le$\,f\,$\le$\,1000\,$\mu$Hz (all the signals, top
panel) or 440\,$\le$\,f\,$\le$\,500\,$\mu$Hz (only high-power signals, bottom
panel).
The power is normalized to the peak at 10.52 (10.06) $\mu$Hz, which
is partially due to the one day aliases (see the text).
Right panel: autocorrelation function of the data DFT using the same subset
as in the left bottom panel. The peaks between about 8 and 13 $\mu$Hz are
partially due to the spectral window, as it is highlighted by the
autocorrelation of the spectral window (dotted line).}
\end{figure*}

\subsection{Period spacing and mode trapping}

The two peaks in the amplitude spectrum DFT, described in the previous
section, could also be due to the period spacing between modes with
successive overtones.
In Figure 7 (left panels) we show the DFT of the period spectrum
(amplitude spectrum in the period domain).
For the upper panel we used a subset of the period spectrum
with periods between 1000 and 2857\,s, while for the lower panel we used
a narrower part with periods spanning 2000\,--\,2273\,s.
Excluding the peak at about 46\,s, which is related to the one day alias
as discussed in the previous section, the most significant period spacings
are 18.8\,s and 10.4\,s (at least in the lower graph; in the upper graph
the 10.4\,s peak appears more uncertain).
Their ratio is close (accuracy better than 5\%) to the asymptotic value of 
$\sqrt{3}$, suggesting that 18.8 and 10.4\,s might correspond to the
l=1 and l=2 period spacings.
In this hypothesis, the differences between the two left panels of Figure 7
suggest that the l=2 modes might be present only (or mainly) in the
high-amplitude region between 2000 and 2273\,s.

An attempt to confirm the hypothesis that the modes of HS\,2324 are equally
spaced in period (and not in frequency) has been done applying the
Kolmogorov-Smirnov (K-S) test (Kawaler 1988) and the Inverse Variance
technique (O'Donoghue 1994) to the first 12 periods listed in Table 2
(excluding the linear combinations).
The results, reported in Figure 7 (right panels), do not confirm that the
modes are equally spaced in period.
Moreover, the lack of any significant period spacing further indicates
that the period list is not complete
\footnote{Note that these methods, which are in general much more reliable
than the DFT of the period spectrum, can completely fail when they are applied
to a period list erroneous and/or incomplete, as it can be in our case.
For this reason the hypothesis that the signals are equally spaced in period
(and not in frequency) can not be completely ruled out.}
\hspace{-1.5mm}
.

Nothing may be said about the trapped modes phenomenon apart the following.
The ratio between the frequencies of the highest peaks in the 380 and 475
$\mu$Hz regions gives $\sqrt{3}$/2 with an accuracy better than 1$\%$.
This number was found in other GW Vir stars, as RXJ\,2117+3412 and the central
star of NGC\,1501 (Bond et al. 1996), and is compatible with calculated
trapping coefficients (Kawaler \& Bradley 1994).

\begin{figure*}[htb]
\vspace{88mm}
\includegraphics{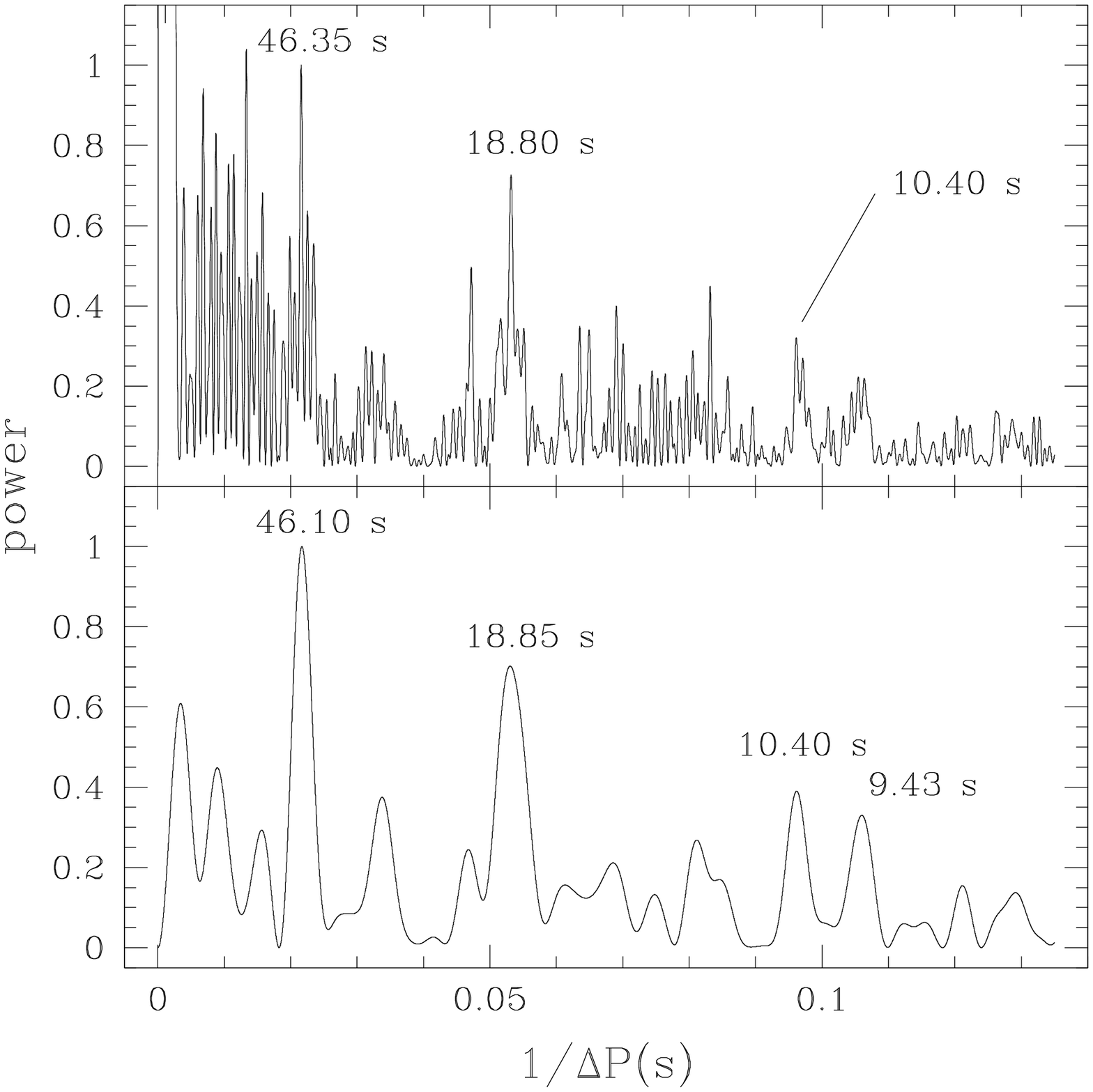}
\includegraphics{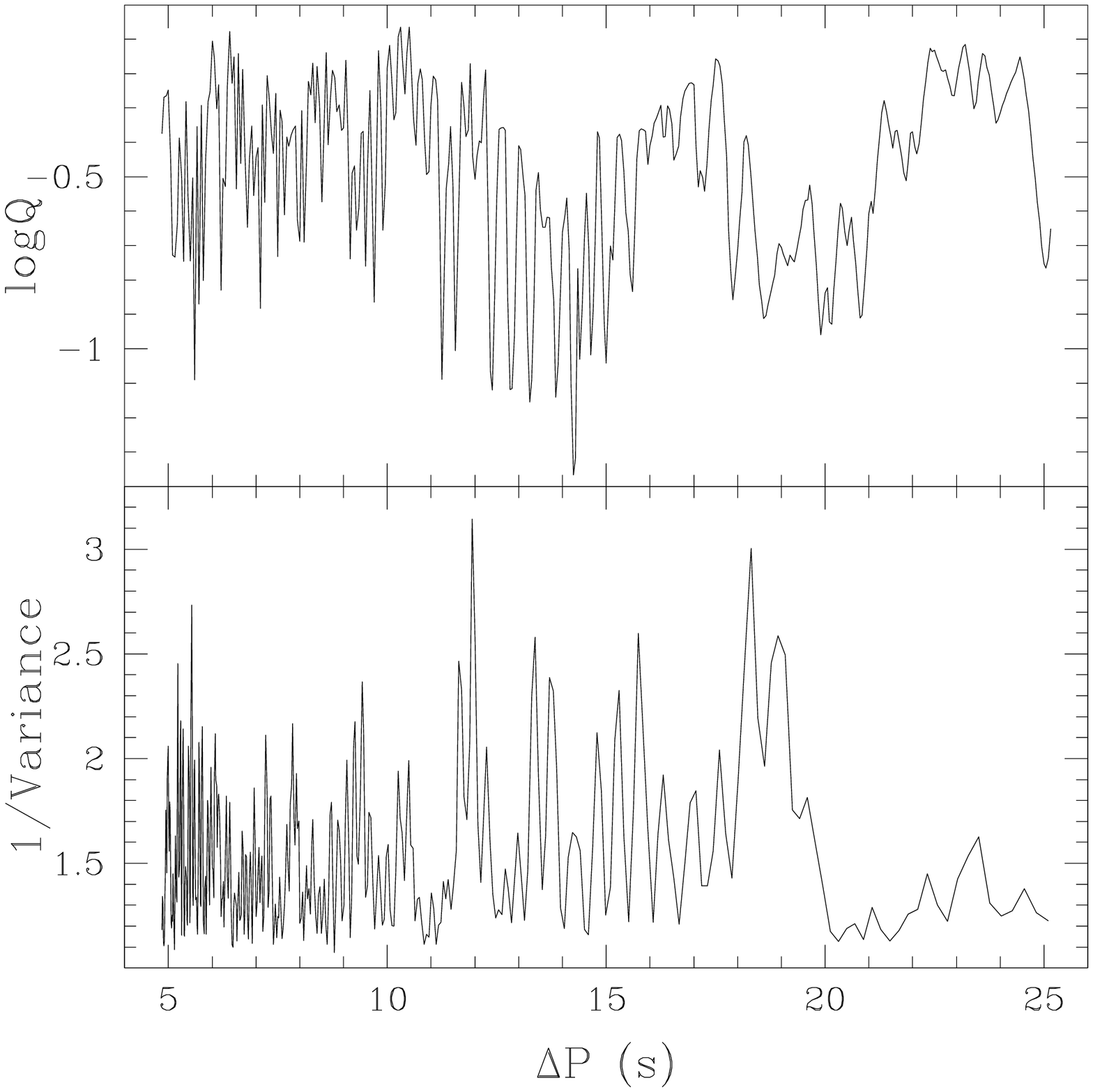}
\caption[]{Search for the period spacing.\\
Left panels: Fourier transform of the period spectrum (amplitude spectrum in
the period domain) in the period range 1000\,$\le$\,P\,$\le$\,2857\,s (top) and
2000\,$\le$\,P\,$\le$\,2273\,s (bottom).
The power is normalized to the peak at 46.3 (46.1)\,s, which is mainly due
to the one day aliases in the main power region between 2000 and 2273\,s.
The peaks at 18.8 and 10.4\,s might be due to the l=1 and l=2 period spacing.
The peak at 9.4\,s is the first harmonic of the 18.8\,s signal.
Right panels: Kolmogorov-Smirnov test (top) and Inverse Variance test (bottom)
applied to the first 12 frequencies listed in Table 2 (excluding the linear
combination region). Both tests do not show any significant value for the
period spacing.}
\end{figure*}

\section{The damped oscillator hypothesis}

When we discovered that the DFT was unstable, we also tried to explain
such apparent time dependence of the DFT with a completely different
quasi-periodic approach. We considered the hypothesis that the DFT
temporal instability was real and due to a very short life time of the
oscillations, which were continuously excited and damped. 

We therefore applied to the HS\,2324 data the Linear State Space model
developed by Michael K\"onig for the analysis of X-ray variability of AGN
(K\"onig \& Timmer 1997, K\"onig et al. 1997).
The current version of this program requires uninterrupted and equally spaced 
datasets. Moreover, in order to provide reliable results, the time scales to be
investigated must be sampled at least ten times.
The only part of our light curve which fulfills these criteria
(JD 94\,--\,94.9, after rebinning with 200\,s) can be actually fitted
with a period of 2134\,s and a damping time of approximately 3.5 periods.
A further attempt has been done using a larger nearly uninterrupted part
of the light curve (JD 94\,--\,95.7), filling the small gaps with white noise
or with synthetic data (both techniques give same results).
The results are slightly different in this case: 2154\,s and 3.1 periods.
In both cases from a K-S test the residual is white noise with over 90$\%$
probability.
If we try to find a secondary period the results are unreliable
(damping time longer than the dataset), but in any case the inclusion of more
frequencies does not improve the fit.

In Figure 8 the fit from the Linear State Space model is compared with the
multisinusoidal fit: the quality is comparably good.
Despite this partial success, we cannot demonstrate that the damping time
found is really a fundamental quantity, constant over at least some days.
We would need several datasets (ideally, but not necessarily coherent) of at
least one day length to reject or corroborate this hypothesis.
Moreover, the excitation time-scale obtained from the Linear State Space model
appears to be very short respect to the growth rates obtained from GW Vir
non-adiabatic models.
For these reasons the present results are not convincing enough to abandon
the DFT results.
On the other hand, the inviting advantage of the quasi-periodic approach
is the small number of parameters required to describe the light curve.
Unstable power spectra have been found also in other luminous PG\,1159
stars (e.g. RXJ\,2117+3412) and [WC] CSPN (e.g. NGC\,1501) variables.
Changes were observed down to the time resolution of several days
(Bond et al. 1996, Table\,6).

\begin{figure*}[htb]
\vspace{118mm}
\includegraphics{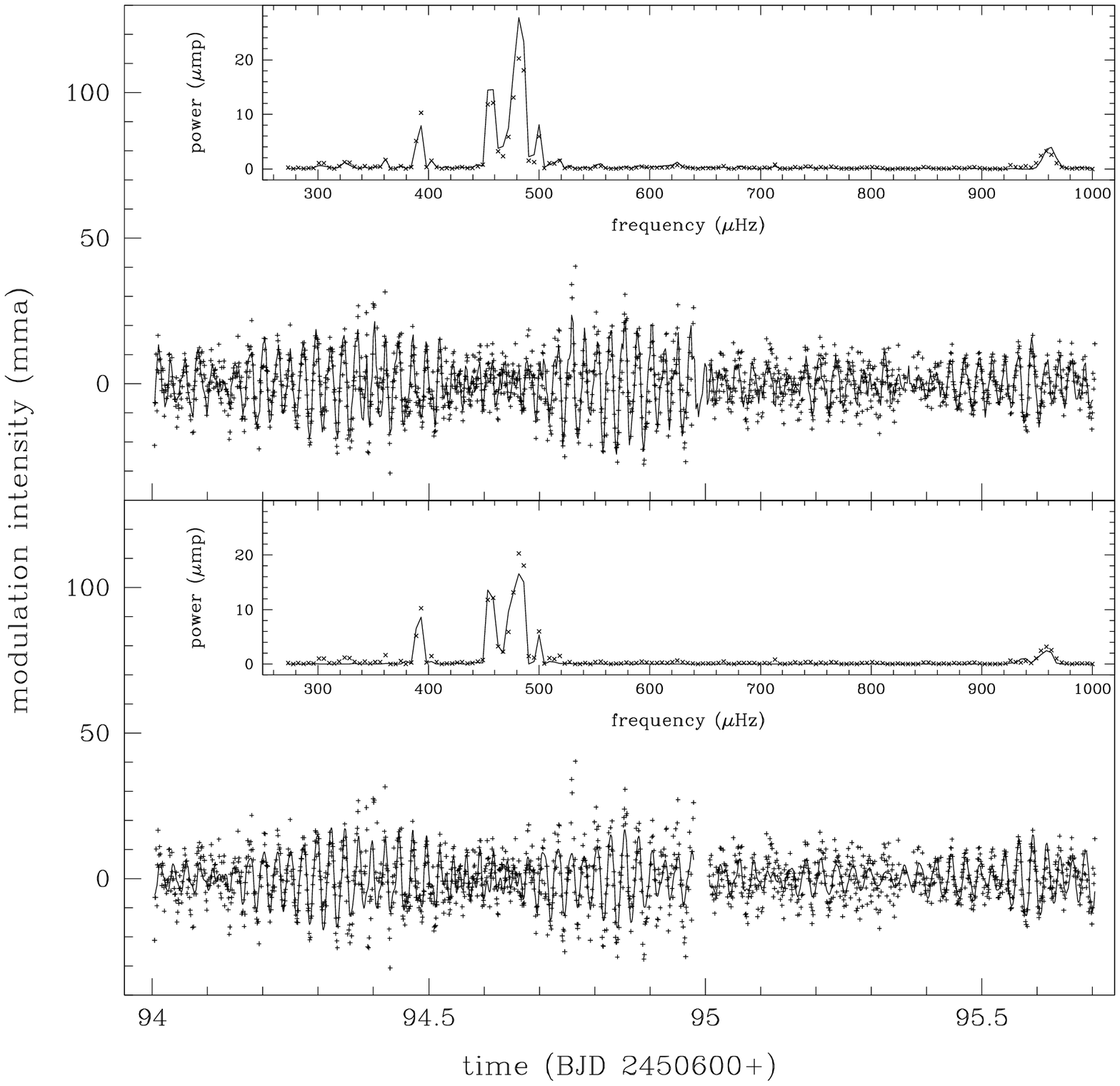}
\caption[]{Damped oscillator (up) vs 19 sinusoids function (down) in the
best part of the light curve (BJD 94.0\,--\,95.7).
Both the synthetic light curves (main panels) and DFTs (window panels)
are shown and compared with the HS\,2324 data (crosses).}
\end{figure*}

\section{Summary and discussion}

The results of our multisite campaign clearly show that the power spectrum
of HS\,2324 contains several periodic signals (about 20 or more).
We therefore may exclude binarity to explain its variability, as already
suggested by Handler et al. (1997).
The frequencies and amplitudes are comparable with those of the GW Vir stars.
If we consider also that HS\,2324 is spectroscopically classified
as a PG\,1159 star, the immediate interpretation is that its variability is
due to high overtone g-mode pulsations.

The high H abundance detected in HS\,2324, about 17\% in mass (Dreizler 1998),
is a very interesting and unique (up to now) element, which might help to
shed light upon the driving mechanisms of the GW\,Vir stars.
As discussed in the introduction, the presence of H was generally
considered as an inhibitor of pulsations (Stanghellini et al. 1991).
But this result was rather speculative because, until a few years ago,
no H-rich PG\,1159 stars were known.
With HS\,2324 this question has gained importance.
Presently the real effects of the presence of H appear to be less severe
from preliminary models of H-rich GW Vir stars (Saio 1996, Gautschy 1997).
On the other hand, the detection of hydrogen in the atmosphere of HS\,2324 does
not necessarily imply that hydrogen is present also in the driving regions.
However, HS\,2324 belongs to the subclass of luminous PG\,1159 stars which
still show mass loss effects in strong UV/FUV lines (Koesterke \& Werner 1998,
Koesterke et al. 1998).
It is highly probable that HS\,2324 is also affected by mass loss which would
inhibit an abundance gradient due to gravitational settling.
It is therefore plausible that the atmospheric composition is also
representative for the driving region.
An unambiguous detection of mass loss and the determination of the mass loss
rate has to await FUSE (Far Ultraviolet Spectroscopic Explorer) observations
of the O\,{\sc VI} resonance lines.

However, the asteroseismological analysis is hampered by the fact that the DFT
appears to be unstable in time.
In principle this is not a new phenomenon: it has been observed in the light
curve of all luminous PG\,1159 and [WC] variables (the term "variable
variables" was therefore coined by S. D. Kawaler).
But in our case, as discussed in Section 3.1, this fact is probably due
to a poor frequency resolution caused by an insufficient coverage.

If this interpretation is correct, it is not possible to obtain definite
precision asteroseismology results from our data set.
Nevertheless a spacing of the signals is probably present in the DFT and
can be explained in two different ways.
The most likely hypothesis is that we see the frequency spacing produced by
the stellar rotation with a period $P_{ROT}=2.31\pm0.15$ days.
The second possibility, which can not be completely excluded, is that we see
the period spacing between successive overtones.
In this case the period spacings, equal to 18.8 (l=1) and 10.4 (l=2) s,
would imply a stellar mass of 0.67 (l=1) and 0.70 $M_{\odot}$ (l=2) 
using the interpolation formula of Winget et al. (1991).
This asteroseismological mass would be higher than the 0.59 $M_{\odot}$ value,
found from spectroscopy plus evolutionary tracks (Dreizler et al. 1996).
But this discrepancy would not be very significant as it is possible that
the interpolation formula of Winget et al. (1991) needs some adjustment
because of the peculiar composition of HS\,2324.

In an alternative interpretation we assumed that the DFT instability was real
and we applied the Linear State Space model (K\"onig \& Timmer 1997) to
investigate the quasi-periodic nature of these variations. As discussed in
Section 5, this approach is also partially successful, but it also requires
new longer observations to be confirmed.
At the moment we can only speculate about the physical interpretation.
Do we see the coupling time between different g-modes or the damping of
a single mode?
In principle, the quasi-periodic nature of HS\,2324 could even
endanger the interpretation as g-mode pulsations.

In conclusion both possible interpretations of the apparent DFT
instability need a new bigger observational effort, which could be realized
only with a larger number of telescopes in a WET-like campaign.
If we adopt the hypothesis that the DFT instability is only apparent, it is
also possible to estimate the duration needed for such a campaign in order
to be able to separate all the frequencies.
If all the l=1 and l=2 frequencies were excited in the region between
450 and 500 $\mu$Hz, where most power is concentrated, the average frequency
separation would be about 0.4 $\mu$Hz.
In a more realistic case, if only 50\% of the frequencies were excited
(as in PG\,1159, which is the GW Vir star with the largest number of detected
modes), a frequency resolution of about 0.8 $\mu$Hz would be enough.
Therefore we would require a data set with a time base of about 1.7 times
the data set analyzed in this paper.

\begin{acknowledgements}

This research was partially supported by the Italian ``Ministero per 
l'Universit\`a e la Ricerca Scientifica e Tecnologica'' (MURST) and by the
EU grant ERBFMMACT980343;
by the Deutsche Forschungs Gemeinschaft under travel grant DR 281/8-1;
by the Austrian Fonds zur F\"orderung der wissenschaftlichen Forschung under
grant S7304-AST; and by the Chinese National Science Foundation.
S.D. would like to thank Ralf Geckeler (University of T\"ubingen) for
providing his excellent CCD-photometry analysis program package
as well as Katja Potschmidt (University of T\"ubingen) for her introduction
to the time series analysis package of M. K\"onig and for her fruitful
discussions.
R.S. would like to thank Steven Kawaler (Iowa State University, Ames),
Thomas Strauss (Astronomical Observatory of Capodimonte, Naples),
Pawel Moskalik (Copernicus Astronomy Center, Warsow) and Scot Kleinman
(University of Texas at Austin) for interesting discussions, and
Adalberto Piccioni (University of Bologna) for the availability of
the Loiano photometer.
R.S. and S.D. are grateful to the Vienna Delta Scuti Network Group for
their kind hospitality during the short stay in April--May 1998, 
in which part of this work has been done.
The same applies to G.H.'s stay at the University of T\"ubingen in December
1997.

\end{acknowledgements}


\begin{thebibliography}{}

\bibitem{} Bond H.E., Kawaler S.D., Ciardullo R., et al., 1996, AJ 112, 2699
\bibitem{} Bradley P.A., 1998, Baltic Astronomy vol.7, 355
\bibitem{} Bradley P.A., Dziembowski W.A., 1996, ApJ 462, 376
\bibitem{} Ciardullo R., Bond H.E., 1996, AJ 111, 2332
\bibitem{} Deeming T.J., 1975, Ap\&SS 36, 137
\bibitem{} Dreizler S., 1998, Baltic Astronomy vol.7, 77
\bibitem{} Dreizler S., Werner K., Heber U., Engels D., 1996, A\&A 309, 820
\bibitem{} Dreizler S., Heber U., 1998, A\&A 334, 618
\bibitem{} Gautschy A., 1997, A\&A, 320, 811
\bibitem{} Handler G., Kanaan A., Montgomery M.H., 1997, A\&A 326, 692
\bibitem{} Kawaler S.D., 1988, in Advances in Helio and Asteroseismology,
 Proc. IAU Symp. 123, eds. J.J. Christensen-Dalsgaard and S. Frandsen
 (Reidel:Dordrecht), 329
\bibitem{} Kawaler S.D., Bradley P.A., 1994, ApJ 427, 415
\bibitem{} Koesterke L., Dreizler S., Rauch T., 1998, A\&A 330, 1041
\bibitem{} Koesterke L., Werner K., 1998, ApJ 500, L55
\bibitem{} K\"onig M., Timmer J., 1997, A\&AS 124, 589
\bibitem{} K\"onig M., Staubert R., Wilms J., 1997, A\&A 326, L25
\bibitem{} Kurtz D.W., 1985, MNRAS 213, 773
\bibitem{} Loumos  G.L., Deeming T.J., 1978, Ap\&SS 56, 285
\bibitem{} Moskalik, P., 1993, Baltic Astronomy vol.2, 485
\bibitem{} Nather R.E., Winget D.E., Clemens J.C., Hansen C.J., Hine B.P.,
 1990, ApJ 361, 309
\bibitem{} Napiwotzki R., Sch\"onberner D., 1991, A\&A 249, L16
\bibitem{} O'Donoghue D., 1994, MNRAS 270, 222
\bibitem{} Pesnell W.D., 1985, ApJ 292, 238
\bibitem[1992]{PTVF:92} Press W.H., Teukolsky S.A., Vetterling W.T.,
 Flannery B.P., 1992, Numerical Recipes, (Cambridge Univ. Press: Cambridge)
\bibitem{} Saio H., 1996, in Hydrogen-Deficient Stars, eds. U.Heber and
 C.S.Jeffery, ASP Conf. Series 96, 361
\bibitem{} Silvotti R., 1996, A\&A 309, L23
\bibitem{} Stanghellini L., Cox A.N., Starrfield S., 1991, ApJ 383, 766
\bibitem{} Starrfield S., Cox A.N., Kidman R.B., Pesnell W.D., 1984, ApJ 281,
 800
\bibitem{} Stumpff P., 1980, A\&AS 41, 1
\bibitem{} Werner K., 1992, in Atmospheres of Early-type Stars, eds. U. Heber
 and C.S. Jeffery, Lecture Notes in Physics 401, (Springer-Verlag: Heidelberg),
 273
\bibitem{} Werner K., Bagschik K., Rauch T., Napiwotzki R., 1997, A\&A 327, 721
\bibitem{} Winget D.E., Nather R.E., Clemens J.C., et al., 1991, ApJ 378, 326
\bibitem{} Winget D.E., Nather R.E., Clemens J.C., et al., 1994, ApJ 430, 839

\end{thebibliography}
\end{document}